\documentclass[useAMS,usenatbib]{mn2e}
\usepackage{graphicx}

\title[Disky cluster red sequence galaxies]{Morphological evolution in situ: Disk-dominated  cluster red sequences  at $z\sim 1.25$}
\author[R. De Propris et al.]
{Roberto De Propris$^{1}$\thanks{E-mail:email@address (AVR); otheremail@otheraddress (ANO)} 
Malcolm N. Bremer$^{2}$ and Steven Phillipps$^{2}$\\
$^{1}$ Finnish Centre for Astronomy with ESO, University of Turku, Finland\\
$^{2}$ H.H. Wills Physics Laboratory, University of Bristol, Tyndall Avenue, Bristol, BS8 1TL, UK, United Kingdom}
\begin{document}

\date{}

\pagerange{\pageref{firstpage}--\pageref{lastpage}} \pubyear{2002}

\maketitle

\label{firstpage}

\begin{abstract}
We have carried out a joint photometric and structural analysis of red sequence galaxies in four clusters 
at a mean redshift of $<z> \sim 1.25$ using optical and near-IR HST imaging reaching to at least 3 magnitudes 
fainter than $M^*$. As expected, the photometry and overall galaxy sizes imply purely passive evolution of stellar 
populations in red sequence cluster galaxies. However, the  morphologies of red sequence cluster galaxies at these 
redshifts show significant differences to those of local counterparts. Apart from the most massive galaxies, the high 
redshift red sequence galaxies are significantly diskier than their low redshift analogues. These galaxies also show 
significant colour gradients, again not present in their low redshift equivalents, most straightforwardly explained by 
radial age gradients. A clear implication of these findings is that red sequence cluster galaxies originally arrive on the 
sequence as disk-dominated  galaxies  whose disks subsequently fade or evolve secularly to end up as high S\'ersic index 
early-type galaxies (classical S0s or possibly ellipticals) at lower redshift. The apparent lack of growth seen in a comparison 
of high and low redshift red sequence galaxies implies that any evolution is internal and is unlikely to involve significant 
mergers. While significant star formation may have ended at high redshift, the cluster red sequence population continues to
evolve (morphologically) for several Gyrs  thereafter.

\end{abstract}


\section{Introduction}

The common picture of galaxy populations in clusters is that they are remarkably homogeneous in nearby 
as well as distant objects, out to $z \sim 1.5$ at least, if not beyond. The locally measured luminosity functions
and colour-magnitude relations show little or no significant variation as a function of gross cluster properties
\citep{DePropris2003a,Lopez2004}. There is no clear evidence for evolution in the typical  near-IR luminosity (and 
therefore stellar mass) of galaxies to $z \sim 1.5$, other than pure passive evolution of their stellar populations 
\citep{DePropris1999,DePropris2007,Andreon2006,Andreon2008,Muzzin2008,Mancone2010,Mancone2012} at $\sim L^*$ 
and brighter, implying no significant growth of cluster galaxies over the past 2/3 of the Hubble time. Similarly, distant 
clusters studied thus far are usually dominated by red sequence galaxies with colours consistent with passively evolved 
local samples, having high formation redshifts and short ($\sim 1\,$Gyr) star formation timescales \citep{Kodama1997,
Blakeslee2003,Mei2006a,Mei2006b,Mei2009,Mei2012}. This argues for a model where cluster galaxies assembled most of
their stellar content at early times and with their stellar populations formed over short periods \citep[e.g.][]{Pipino2004}. 

Unlike old early-type field galaxies at similar redshifts \citep{Longhetti2007,Cimatti2008, Damjanov2009,Whitaker2012,
Cassata2013,Williams2014} which are  more compact than  similarly massive low redshift counterparts \citep{Shen2003}, 
the sizes of red cluster galaxies also do not appear to have evolved appreciably \citep{Cerulo2014,Delaye2014,Jorgensen2013,
Jorgensen2014} since high redshift. Consequently, the red sequence cluster galaxies have been presumed to be "red and dead", 
at least as far as their star formation activity and growth is concerned.

More recent work has shown that at least for some clusters at higher redshifts  there is evidence for a population 
of massive blue star-forming galaxies in their centres \citep{Brodwin2013,Zeimann2013,Alberts2014,Mei2014} with a 
possible 'reversal of fortune' in the morphology-colour-density relation at $z > 1.5$ \citep{Tran2010}. This is unlike the 
classical Butcher-Oemler effect \citep{Butcher1978,Butcher1984,Dressler1984} where the excess blue systems in high
redshift clusters tend to be low mass galaxies \citep{DePropris2003b,Pimbblet2012}, but instead involves  star-forming 
galaxies with masses comparable to the spheroids  that dominate the red sequence population in the local Universe. At 
these redshifts, there appears to be diversity in star formation and growth between clusters, e.g. some show evidence of 
on-going mass assembly among luminous galaxies \citep{Rudnick2012,Fassbender2014} while others appear to resemble 
passive local systems \citep{Andreon2014,Koyama2014}. 

This diversity is the likely signature of the end of the epoch of significant star formation and growth in the cores of massive clusters, 
any later evolution in this galaxy population is necessarily more  subtle. Morphologically, it is usually assumed that passive red colours 
correlate with spheroid dominated structures, i.e. the red sequence galaxies are early types, but we should note that passive or red spirals 
(disks) also exist at low redshifts \citep{Koopmann1998, Wolf2009} which indicates that stellar population and structural evolution do not 
have to be synchronous.
 
One key advantage of studying cluster galaxy populations as opposed to field or group systems is that once a galaxy resides within 
a massive cluster it remains in such an environment thereafter. We can therefore identify the progenitors of local red sequence galaxies 
in distant clusters precisely because of the lack of significant evolution in luminosity functions and colours, whereas in the field galaxy 
population evolution can only be measured through the statistical changes in the properties of the sample at different redshifts. The aim
of this paper is to measure the evolution of luminosity, colour, size and shape for galaxies  in clusters at $1.0 < z < 1.4$ to provide an
overall view of galaxy evolution in the densest environments. This redshift interval may correspond to the highest lookback time at which 
the more massive cluster galaxies  are still  quiescent and a direct line of descent can therefore be drawn to local samples. The properties of 
cluster galaxies may then offer clues to the more general problem of galaxy evolution, in the same fashion as stars in clusters have been 
instrumental (through their homogeneity) to our understanding of stellar evolution. As we will show, while the red sequence at $z<1.4$ 
may well be populated by galaxies that have finished forming stars, and assembling their mass, they must still be evolving significantly in 
their morphological properties for a significant period of time thereafter.

\section{Dataset}

In order to probe both the photometry and morphology of red sequence galaxies down to low near-IR luminosities (and consequently 
stellar masses) at $1 < z < 1.4$ we require HST data of sufficient depth in both optical and near-IR bands. The infrared is needed to 
select galaxies as closely as possibly by stellar mass and explore its 2D distribution. Ground-based data simply cannot reach the fluxes 
and surface brightnesses necessary at these redshifts. Although a significant number of the known $z>1$ clusters have been imaged 
by HST and therefore have data in the archive, currently only a few have sufficiently deep data in a sufficient range of bands. For 
example, we do not consider the two Lynx clusters at $z=1.27$ as only optical data (but no infrared) are available in the archive. 

For this work we aim to be able to reach at  least M$^* + 2.5$ in the observed near-IR, which is roughly equivalent to a stellar mass  
comparable to that of the LMC (so ${\rm log}_{10}(M_{stellar}/M_\odot) \sim 9.5$) assuming only passive evolution in luminosity. 
We then require optical data (probing the rest frame blue/near-UV) of a depth sufficient to identify and characterise most or all of the 
red sequence galaxies in combination with the near-IR. In practice, the reddest  band available in the archival data is the WFC3
$H_{F160W}$, which corresponds to the rest-frame $R$ band at the redshifts of interest here.  Given the current state of the archive 
we chose to study four clusters at $1 < z < 1.4$ with data  of sufficient quality and depth. These four systems are XMM1229 at $z=0.98$, 
RDCS1252+2927 at $z=1.24$, ISCS1434 at $z=1.24$ and XMM2235 at $z=1.40$,  with near-IR data obtained as part of Proposal 12051, 
PI Perlmutter. Exposure times were of around 4ks each. Optical imaging in the $i$ and $z$ bands, to sample the rest-frame $U$ and $B$ 
were derived from a variety of sources (XMM1229 from program 10496, PI: Perlmutter; RDCS1252 from PID 9290, PI:  Ford; ISCS1434 also 
from PID 10496 and XMM2235 from PID 10531, PI: Mullis). Exposure times vary between 1.5 and 7.2 ks in $i$ and  between 3 and 10 ks 
in $z$.

Initial photometry was carried out exactly as in our earlier work \citep{DePropris2013}. We used {\tt Sextractor} \citep{Bertin1996} with the same  
parameters as used previously to measure both total and aperture magnitudes on each separate image and bandpass.  In addition to the 
cluster fields, we used similarly-deep reference fields which are treated in exactly the same way in order to understand the expected contribution 
from foreground/background interlopers when determining luminosity  functions. For the reference fields we used the same  data on the 
Extended Groth Strip as that used in \cite{DePropris2013}, as well as $H$ band images of the Early Release Data 2 field, which covers part 
of the southern GOODS field \citep{Giavalisco2004}. The latter provides us with $i$ and $z$ photometry, allowing us to derive the red 
sequence luminosity functions in addition to the overall cluster luminosity functions. Note all magnitudes are on the AB system.
 
\begin{figure}
\includegraphics[width=0.5\textwidth]{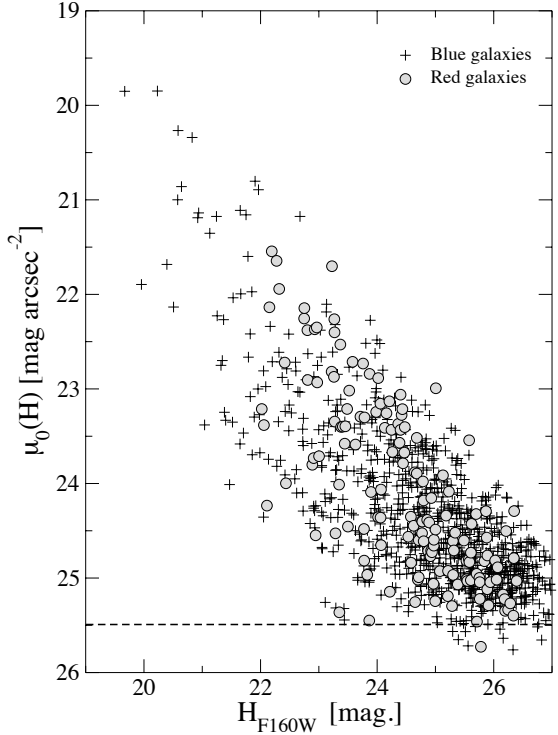}
\caption{Central surface brightness vs. total magnitude in the $H$ band for galaxies in our clusters. 
Red sequence galaxies (defined below) are plotted as filled grey circles, while blue galaxies (including
non cluster members) are shown as plus signs. The diagonal cutoff at bright surface brightnesses 
represents the locus of stars. There is a region below $\mu(H)_0 \sim 25.5$ mag arcsec$^{-2}$ where
the density of galaxies falls at $H > 23.5$, giving the impression of a 'cut' in the plot. This (shown as a
dashed line) is the explicit surface brightness limit of our data. Galaxies may be brighter than the actual 
completeness limit but still be undetected if their central surface brightness is below the sky (cf.
Disney 1976, Phillipps \& Driver 1995).}
\label{subplot}
\end{figure}

For star-galaxy separation, we plot the central surface brightness (measured in the Sextractor detection aperture) vs. total magnitude 
in Fig.~\ref{subplot}. Stars are identified as a high surface brightness `edge' at all luminosities. All objects remaining after star-galaxy 
separation  were visually inspected to remove possible contaminants such as cosmic ray streaks, diffraction spikes, brighter galaxies 
segmented by the algorithm and (for cluster fields) large gravitational arcs. Note that we performed a similar procedure on our reference 
fields as well. These will be discussed in more detail in a future paper.

In addition to the standard flux cuts typically used in photometric selection of galaxy samples, we also consider the effect of surface 
brightness on the completeness of any sample of red sequence objects. While luminous galaxies are detected within a broad wedge-shaped 
region in Fig.~\ref{subplot},  there is a horizontal `cut' at low surface brightnesses, which represents the detectability limit set by the sky 
brightness. While this significantly affects the bulk of the galaxy distribution at $H\sim 25$, comparatively low surface brightness outliers 
start to be lost by  $H > 23.5$, so we limit our sample selection to brighter than this. The surface brightness limit induces a significant selection 
effect (e.g., \citealt{Disney1976,Phillipps1995}) which we will discuss later in this paper  in relation to the red sequence luminosity functions. 

\section{Colour and Luminosity Evolution}

We first reconfirm several results from published studies of the luminosity functions and colour magnitude relations of distant clusters. 

The data at hand reach to as faint \citep[e.g.][]{Strazzullo2010} or fainter luminosities than has been previously achieved at this redshift, 
enabling us to better measure the faint-end slope of the galaxy luminosity function and study the evolution of the red sequence, which allows 
us to relate our further analysis of sizes and morphologies to galaxies more representative of the descendant population in lower redshift clusters 
rather than purely the most massive galaxies. 

As in our previous paper on a similar HST dataset at $0.2 < z < 0.6$ \citep{DePropris2013}, we derive composite luminosity functions in the $H$ 
band from the background-subtracted counts in each cluster field, following the procedure described by \cite{Colless1989}. We have shifted all 
clusters to a mean redshift of $z=1.25$, following the procedure in \cite{DePropris1999}, assuming a distance modulus from the conventional 
cosmological model\footnote{Throughout this paper we assume the  concordance values for cosmological parameters: $\Omega_M=0.27$, 
$\Omega_{\Lambda}=0.73$ and H$_0=73$ km s$^{-1}$ Mpc$^{-1}$} and differential $k+e$ corrections (between the cluster $z$ and $z=1.25$) 
from a \cite{Bruzual2003} model forming its stars at $z=3$ with solar metallicity and an e-folding time of 1 Gyr, which generally yields a good fit to 
the colour-magnitude relations of the red galaxies in distant clusters (e.g., \citealt{Mei2009,Rudnick2012}).

\begin{figure}
\includegraphics[width=0.5\textwidth]{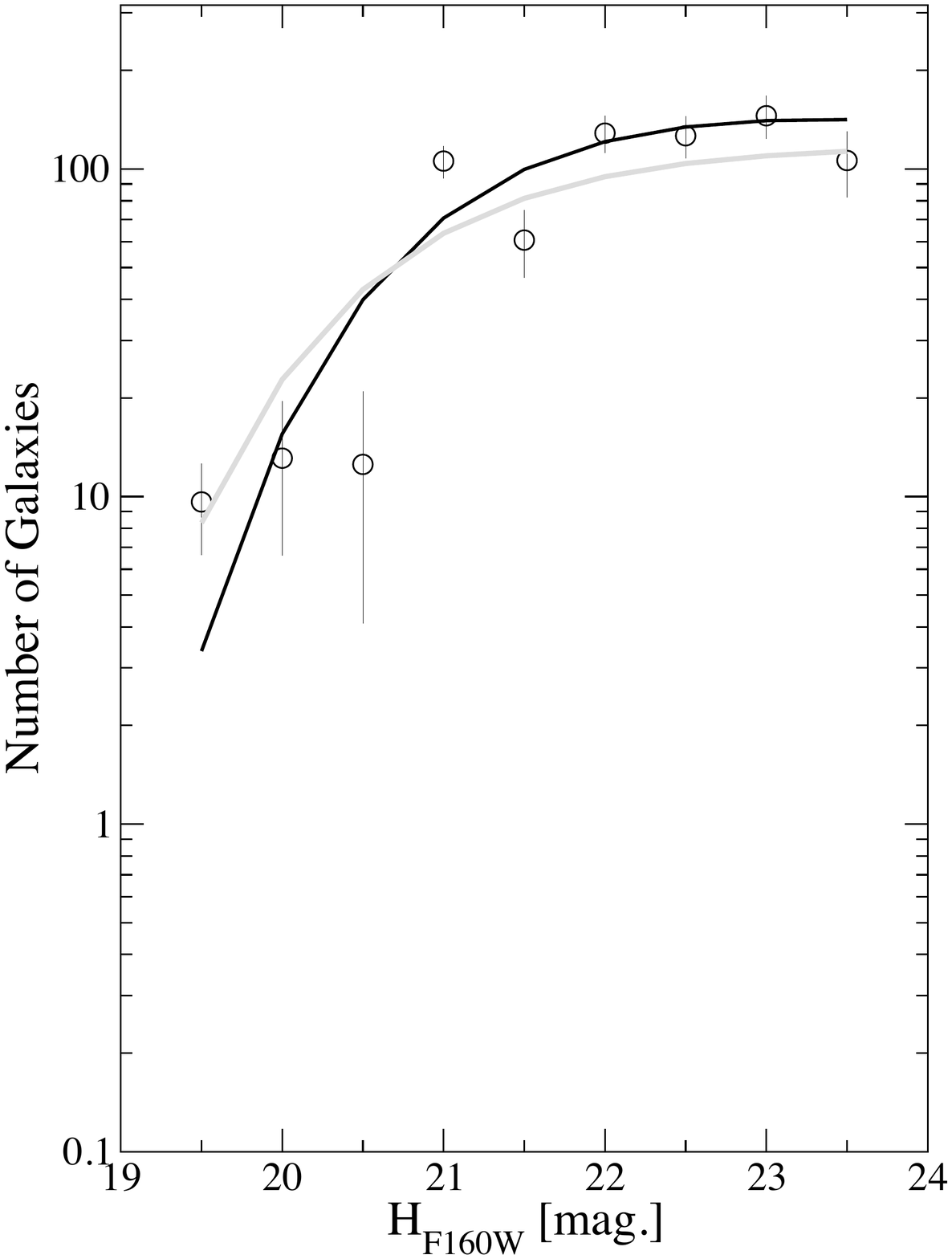}
\begin{picture}(0,0)
\put(100,60){\includegraphics[height=5cm]{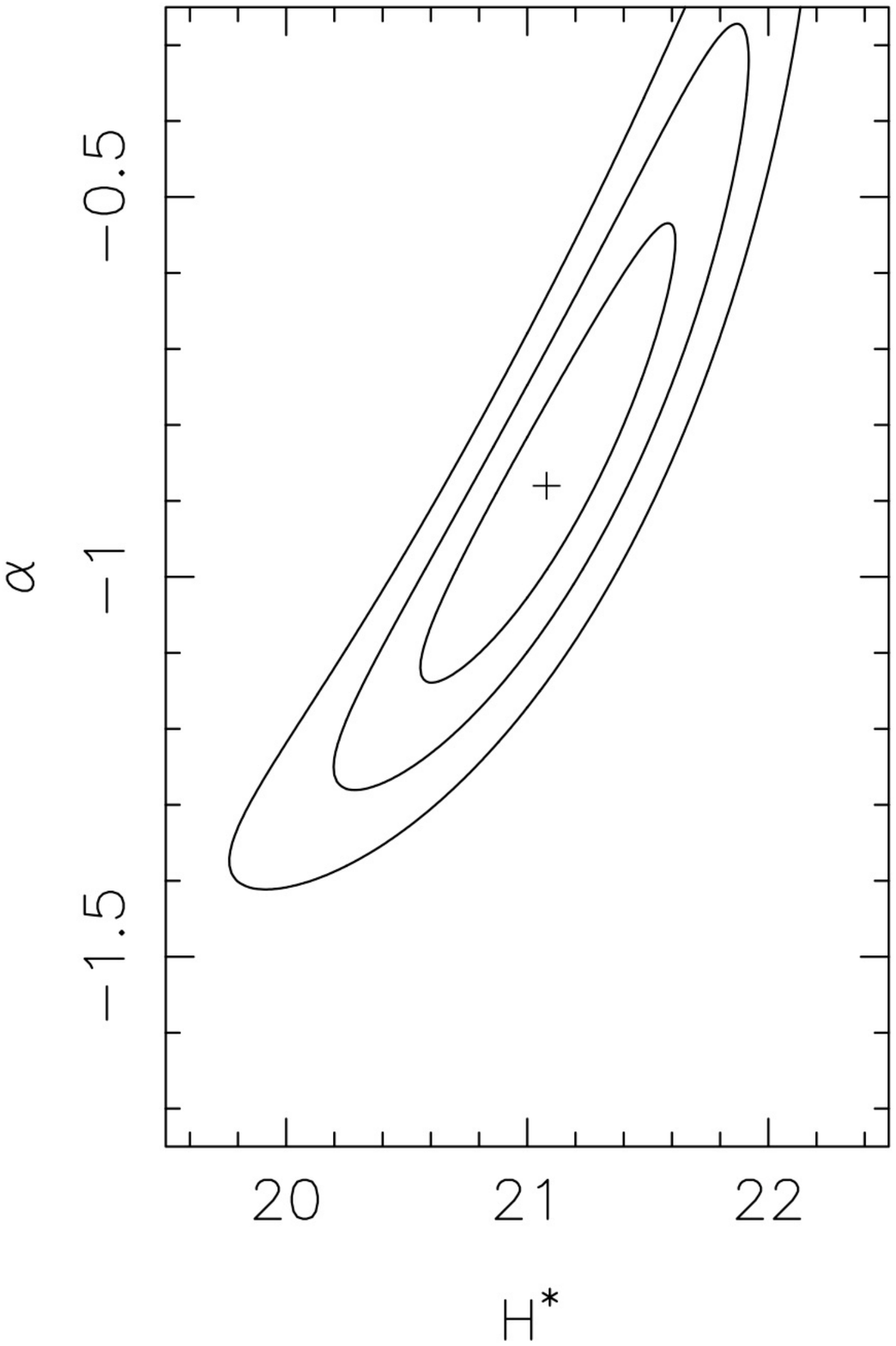}}
\end{picture}
\caption{The composite background-subtracted $H$ band luminosity function for all galaxies in the cluster
fields. The points with error bars show the data, the thick dark line the best fitting Schechter function (with $H^*=21.1$
and $\alpha=-0.78$) and the thick grey line the local cluster $K$-band luminosity function from De Propris \& Christlein (2009),
(having $M_K=-24.5 + 5 \log h$ and $\alpha=-0.98$), assuming $H-K=0.3$ (from Coma) and arbitrary normalisation. We have
adopted the passively evolving model described in the text to shift the local luminosity function to $z=1.25$. The inset shows the $1,2$ and $3\sigma$ 
error ellipses on the values of $M^*$ and $\alpha$, the parameters of the Schechter function. There is no evidence of
evolution (other than purely passive) in the shape of the luminosity function since $z=1.25$ over a range of $\sim100$
in luminosity.}
\label{hlf}
\end{figure}

Fig.~\ref{hlf} shows the composite $H$ band luminosity function derived from the combination of the four clusters, having shifted the 
photometry of each to $z=1.25$, the best fitting Schechter function and the derived error ellipses. Errors include a contribution from 
non-Poissonian clustering computed as in \cite{Huang1997} and \cite{Driver2003}. The best fit to this composite function has $H^*=21.1$ 
and $\alpha=-0.9$ where  $H^*$ is the apparent magnitude at $z=1.25$ corresponding to M$^*$.  The uncertainties on both are shown
in the inset of Fig~\ref{hlf} as $1,2$ and $3\sigma$  error ellipses.  In the figure we also show the composite infrared luminosity function of 
10 local clusters measured from a pure spectroscopic sample in \cite{DePropris2009}, shifted to $z=1.25$ assuming the above passively 
evolving model.

Looking only at the points fainter than $H=22$, the faint end slope appears well-determined (by eye it is consistent with $\alpha \sim -1$), 
but as is usual the uncertainty in M$^*$ and the resulting uncertainty in positioning the exponential cutoff affects the determination of the faint 
end slope. The uncertainty in the bright end is dominated by the background subtraction. As we  show later, the effect of this is decreased 
significantly in the determination of the luminosity function of the red sequence alone, purely because the bright red sequence galaxies fall 
in an otherwise less populated region of colour-magnitude space. A larger cluster sample would be needed to pinpoint $M^*$ more
accurately.

For our chosen cosmology our luminosity function at $z=1.25$ has $M^*_H=-23.6$. Given the local luminosity function from \cite{DePropris2009}
and assuming $H-K=0.3$ for local galaxies, this agrees with the local value of $M^*$ values within 5\%, given the $e+k$ corrections calculated
from the passively evolving model. The implied evolution of $M^*$, in comparison to low redshift clusters, is consistent with a pure passive 
evolution scenario where bright galaxies are assembled at high redshift and suggests that no significant growth has occurred in the galaxy 
luminosities (or  rather, in stellar masses) since at least the redshifts of the present sample of clusters. 

This has already been established by several previous studies (e.g., \citealt{ DePropris1999,Andreon2006,DePropris2007,Muzzin2008,
Mancone2010}), some reaching to even higher redshifts \citep{Andreon2014, Wylezalek2014}. Similarly, the value for $\alpha$ is consistent 
with  the zero redshift values in clusters as measured by \cite{Barkhouse2007} for 57 Abell clusters in $R$. \cite{Andreon2008} and \cite{Mancone2012} 
also found a similar lack of evolution for  $\alpha$ since $z \sim 1$ and 1.3, respectively, while we measured $\alpha \approx -1$ in $I$ for several 
clusters at $0.2 < z < 0.6$ in our previous work \citep{DePropris2013}. 

There is therefore no evidence for significant growth in the luminosity of typical cluster galaxies since $z=1.25$ down to at least 2.5 magnitudes 
below the $M^*$ point. Consistent with the results of \cite{Andreon2014} and \cite{Wylezalek2014}, it appears that any evolution must essentially 
preserve the shape of the galaxy luminosity function across more than a decade in stellar mass. Any mergers and accretion from the field 
would have to be very finely compensated by galaxy growth and destruction to produce remarkably self-similar objects across two thirds 
of the age of the Universe.

\subsection{Red Sequence Luminosity Functions}

\begin{figure*}
\includegraphics[width=\textwidth]{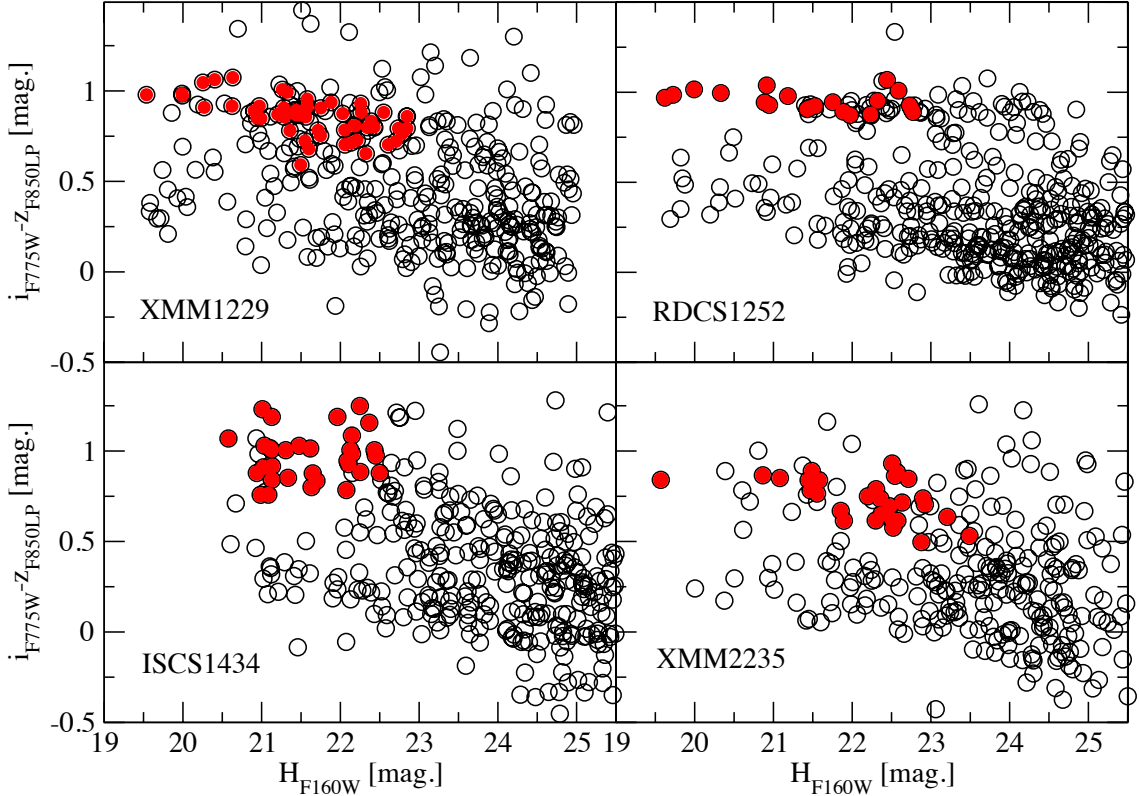} 
\caption{Colour-magnitude diagrams in $i-z$ vs. $H$ for galaxies in all our cluster fields, as identified in 
the figure legend. Red sequences can clearly be seen for all clusters. Galaxies for which we derive
morphological parameters are marked with filled red circles. Failures include galaxies which are on
the edges of the fields, or for which GALFIT only returns highly uncertain parameters.}
\label{hzcmr}
\end{figure*}

In order to obtain the red sequence luminosity function, we must first consider the colour-magnitude diagrams of the clusters. We generate 
colour-magnitude relations for galaxies in the cluster fields in $i-z$ vs. $H$ as shown in  Fig.~\ref{hzcmr}, where colours were measured in 
$0.5''$ apertures and magnitudes are total magnitudes. The existence of well-defined red sequences is apparent, though of varying strength 
between clusters reflecting their overall richness. As expected, their colour-magnitude relations are  consistent with passive evolution of the 
local relations to the appropriate redshift (cf. \citealt{Kodama1997, Mei2012, Snyder2012}). The relations can be followed to the magnitude 
limits of our data, with no apparent evidence for a weakening at the faint end in the best defined cases (e.g. RDCS1252).

We select out red sequence galaxies in the following manner: we derive the slope and intercept of the red sequence by minimum absolute 
deviation \citep{Armstrong1978} and then take all objects within $\pm 0.25$ mag. of the ridge line as belonging to the red sequence. We make 
the same cuts on the colour-magnitude distributions of field galaxies in the ERS2 field to estimate the contribution from foreground and 
background galaxies, normalising to the areas of the observed cluster fields. As with the total luminosity functions in Fig.~\ref{hlf}, we have then 
shifted all clusters to the mean redshift of $z=1.25$ and we have estimated the non-Poissonian contributions to the errors.

\begin{figure}
\includegraphics[width=0.5\textwidth]{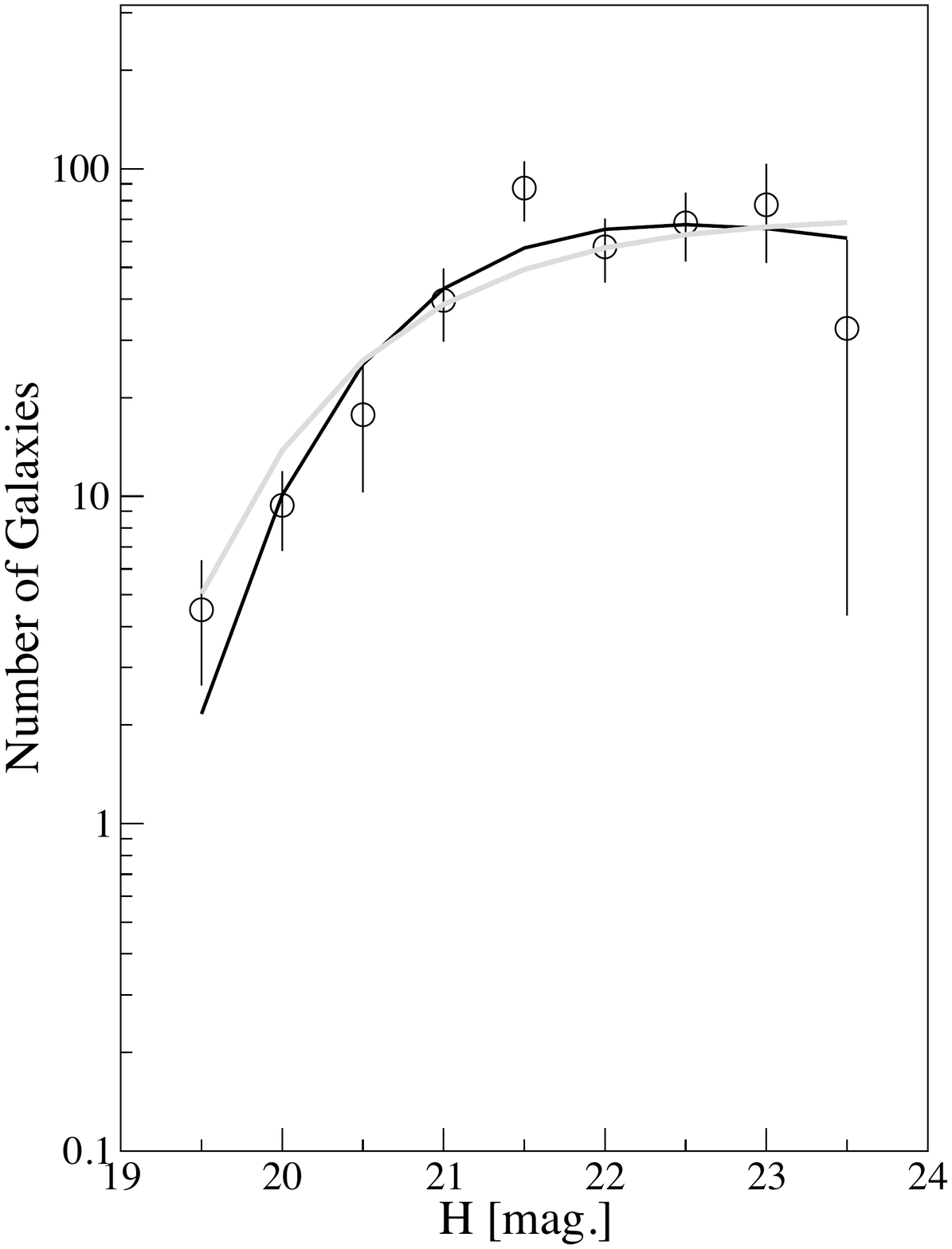}
\begin{picture}(0,0)
\put(100,60){\includegraphics[height=5cm]{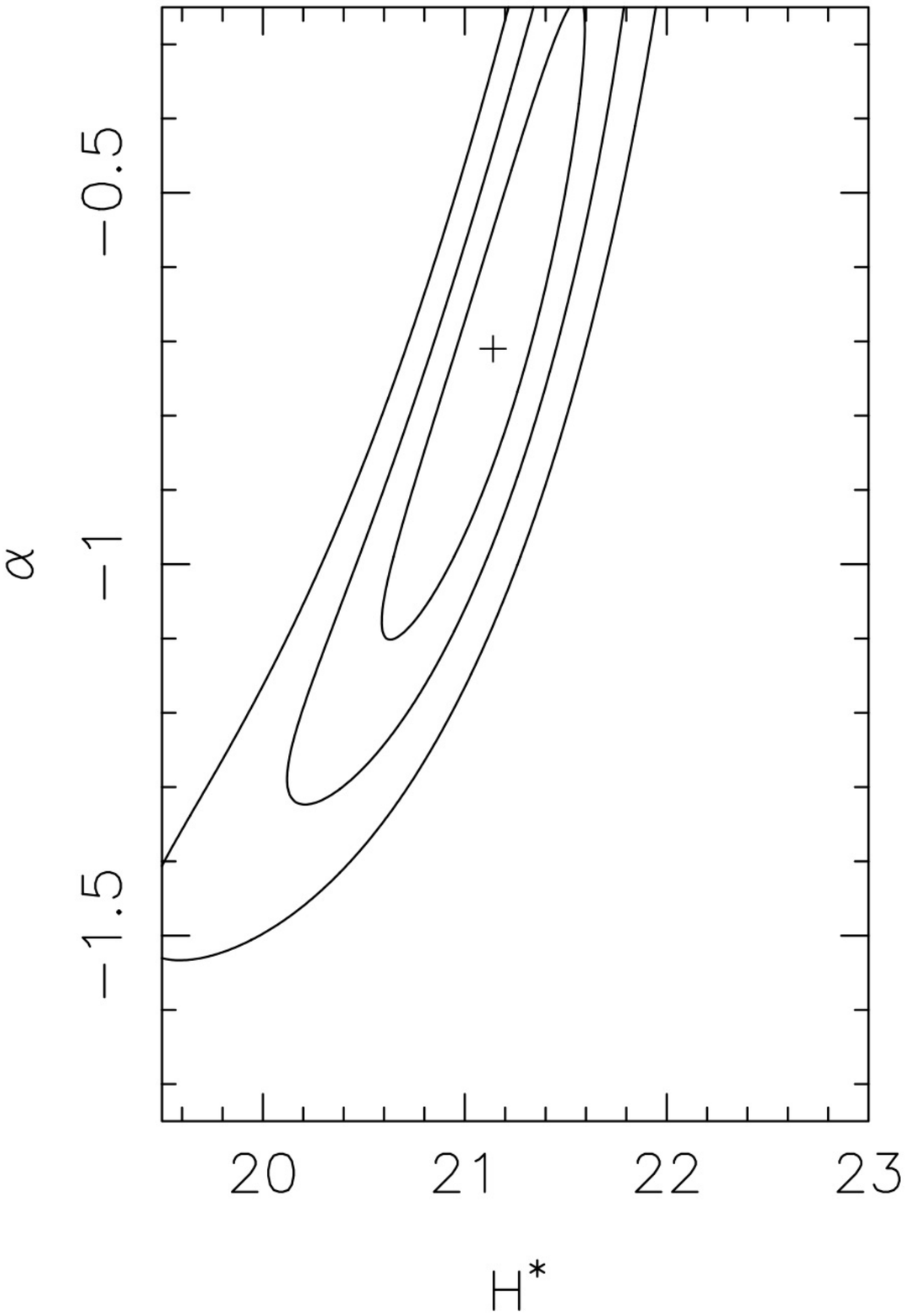}}
\end{picture}
\caption{The composite background-subtracted $H$ band luminosity function for red sequence galaxies in the cluster
fields. The points with error bars show the data, the thick dark line the best fitting Schechter function (with parameters
as in the text) and the thick grey line the local cluster $K$-band luminosity function from De Propris \& Christlein (2009),
with $H-K=0.3$ and arbitrary normalisation, shifted to $z=1.25$ assuming the passively evolving model described in the 
text. The inset shows the $1,2$ and $3 \sigma$ error ellipses on the values of $M^*$ and $\alpha$, the parameters of the Schechter function. 
There is no evidence of evolution (other than purely passive) in the shape of the luminosity function since $z=1.25$ over a
range of $\sim100$ in luminosity.}
\label{hlfred}
\end{figure}

The resulting luminosity function in $H$ for red sequence galaxies is shown in Fig.~\ref{hlfred}. This has $H^*=21.14$ and $\alpha=-0.71$, 
with error ellipses as shown in the inset. We also show the local luminosity function (as in Fig.~\ref{hlf}), shifted to the redshift of the clusters
as described for the previous figure. Even though the local luminosity function is for all galaxies, red sequence galaxies dominate local
clusters. Our measured $H^*$ translates to $M^*_H=-23.6$, again in excellent agreement with the local value assuming a passively evolving
model, implying no significant luminosity (mass) growth.  Note that in the Coma cluster \cite{DePropris1998} measured $M^*_H=-22.58$ and
$\alpha=-0.78$ for a sample of 111 members, all of whom but one lie on the red sequence. The expected $k+e$ correction of $-1.4$ mag. brings
these two values in good agreement, given the photometric errors. The similarity in the $M^*$ and $\alpha$ measured for the whole cluster 
(above) and the red sequence, implies that these high redshift systems are already dominated by red, quiescent galaxies at these early times. 
The error bars on individual points are smaller than those in the full luminosity function (Fig.~\ref{hlf})  because the background/foreground 
contamination is lower in that part of the colour-magnitude space. 

The consistency of the faint end  of the LF for the red sequence down to $H \simeq 24$, with that seen in nearby clusters argues that the 
faint end of the red sequence  in massive clusters, at least down to $\sim M^* +3$, is already in place at the redshifts we study. \cite{Crawford2009} 
and \cite{Rembold2012} also  find no evidence that $\alpha$ at high redshifts is significantly different from the present day value, while \cite{Cerulo2014} 
find no weakening or drop-off at fainter luminosities in the red  sequence for XMM 1229+0151, unlike previous claims (e.g., \citealt{DeLucia2007}).

\subsection{Surface brightness selection effects}

There may be several reasons for the discrepancy between our results and those from other studies that have claimed to see evolution
in the faint-end slope of red sequence galaxies. One possibility is surface brightness selection effects, as we have discussed in \cite{
DePropris2013}.  We plot the central surface brightness vs. total magnitude for red and blue galaxies in the cluster fields for the $H$ band 
data in Fig.~\ref{subplot}. It is clear that there is a quite abrupt surface brightness cut-off at $\mu_0(H) \sim 25.5$ mag arcsec$^{-2}$ 
which then generates an ever increasing magnitude incompleteness even if the detection limit of the data is much fainter. The onset of
this effect lies between $H=23$ and 24. 

In addition, we note that red galaxies are more likely to be affected by this surface brightness limit than blue galaxies. They generally 
have lower central surface brightnesses, probably because they lack bright star-forming regions. In particular, red galaxies seem to 
avoid the high concentration regions of this plot at low luminosities, suggesting that fainter red galaxies are more diffuse and therefore 
less likely to be detected. This is a well known characteristic of dwarfs in nearby clusters and is often used to distinguish them from 
background, higher surface brightness, galaxies (e.g. \citealt{Sandage1985}). 

To at least the $M^*+2.5$ level the data are complete and there is no evidence for a weakening of the red sequence. To fainter limits,
it is uncertain whether there is a real decline in the red sequence or whether this is due to a combination of incompleteness due to surface
brightness and insufficient sensitivity in the bluer bandpasses: this cannot be either proven or disproven with present data, but claims
to this effect should be considered with caution \citep{Martinet2014}.

\section{Morphological Evolution}

Having selected a sample of galaxies representative of the red sequence population at $z=1.25$ over a wide range of luminosities, we can 
now explore their morphological properties and compare these to equivalent properties in a sample of zero redshift red sequence members, 
in this case taken from the Virgo Cluster \citep{Ferrarese2006}.  As the sample of high redshift galaxies is chosen based on a simple colour 
cut and without spectroscopic confirmation, a fraction of this sample will be foreground or background interlopers, either with the same intrinsic 
colour, or scattered into the sample through photometric uncertainties.  Although the contamination rate will be low (or zero) at bright magnitudes
it  increases at fainter levels. We assess this as part of the background subtraction procedure for producing the red sequence luminosity function. 
The interloper fraction at a given magnitude bin is simply the difference between the total number count and the count used in the luminosity function 
divided by the total number count. Brighter than $H=22$  contamination is negligible  and does not influence subsequent results. Between $22<
H<23$  the interloper fraction rises to up to 40\% and therefore is taken into account when interpreting results in this magnitude range.

In order to quantify the morphology of the red sequence galaxies  we  use GALFIT \citep{Peng2002,Peng2010} to simultaneously 
measure their sizes (effective radii) and intensity profile shapes (S\'ersic indices) by assuming a single S\'ersic profile fit. We interactively carry 
out fitting on all galaxies in the $H$ and $z$ band images, using bright stars in the cluster fields as a point spread function reference, inspecting 
the residual images as an indicator of the quality of the fit. We limit this  process to galaxies brighter than $H=23.0$. Fainter than this  the radial 
fits become insufficiently reliable and, as shown earlier,  surface brightness incompleteness selects against larger and more diffuse galaxies 
even though  more compact  galaxies are detected to significantly fainter levels.

\subsection{Size and shape evolution}

\begin{figure*}
\includegraphics[width=\textwidth]{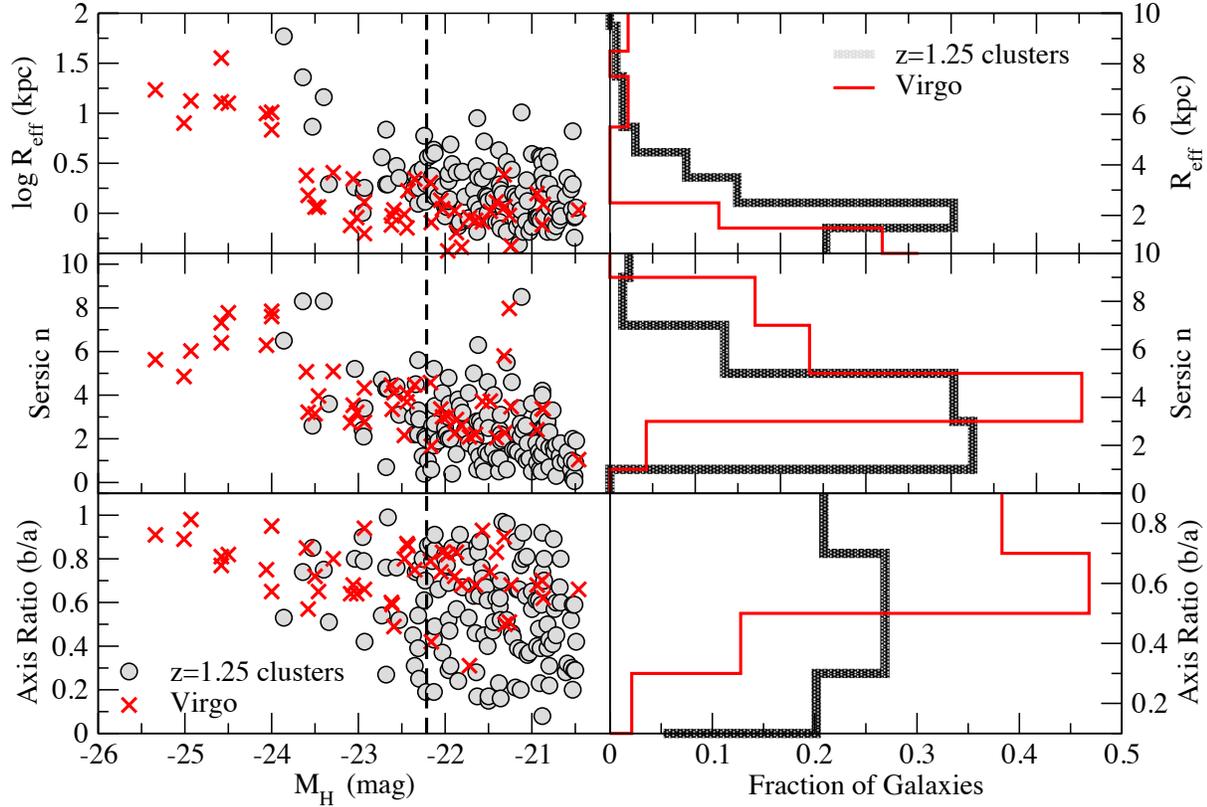}
\caption{{\it Top Panel:} Sizes ($H$-band effective radii) of galaxies in our clusters (see legend); {\it Middle Panel:} S\'ersic
indices (from single S\'ersic fits); {\it Bottom Panel:} Ellipticities (axis ratios), all plotted vs. $M_H$ (evolved to zero redshift) and compared 
to data in the Virgo cluster from Ferrarese et al. (2006). Compare with Fig.6 from Gutierrez et al. (2004): there is
no strong evidence for size evolution. However, there is a population of objects with $n < 2$ in our distant clusters
which is not seen in Virgo galaxies of similar luminosity. These also appear to have more flattened and disk like
shapes as seen in their axis ratios. The dashed vertical line indicates the position of the $M^*$ point. In the equivalent panels on the right
hand side of the figure, we show histograms of the distributions in $R_{eff}$, $n$ and $b/a$ for galaxies in the cluster sample
(grey stippled lines) compared to the counterparts (in luminosity) for Virgo (red lines), demonstrating an excess of galaxies with low $n$ and more elliptical 
axis ratios in the high redshift systems we study.}
\label{figsiz}
\end{figure*}

Fig.~\ref{figsiz} (top panel) shows the effective radii of galaxies in our clusters as a function of $H$-band absolute magnitude for each galaxy 
when evolved to $z=0$ as described earlier.  The equivalent data in \cite{Ferrarese2006} from the ACS Virgo Cluster Survey (VCS) 
is also plotted as the zero redshift comparator. At $z=1.25$,  the observed $H$ band more closely matches the restframe $R$ band, 
while the VCS data are measured in the $z$ band. However, as shown by previous studies (e.g., \citealt{Vader1988,Tamura2000}) and 
the plotted Ferrarese et al. data itself, colour gradients for local ellipticals are small, especially in the redder bands, so we expect that 
there will be little difference between the effective radii in $z$ and $R$ in Virgo. 

We see little evidence of size evolution in cluster galaxies across this redshift range. The high redshift galaxies at M$^*$ and fainter have 
$R_{eff}(H) \sim 1-3$kpc,  comparable to galaxies of similar luminosity (after passive evolution) in Virgo \citep{Ferrarese2006}, Coma and 
Fornax \citep{Gutierrez2004} and typical WINGS clusters \citep{Poggianti2013} at low redshift.  The more luminous galaxies are larger and 
again comparable in size to the corresponding luminosity/stellar mass galaxies in nearby clusters. A K-S test returns that the two distributions
are drawn from the same sample at least at the 90\% level. As with previous work (e.g., \citealt{Delaye2014}) this implies that there is 
significantly less size evolution in red sequence cluster galaxies over the past $\sim 9$Gyr than that seen for field galaxies of similar mass.  

Many previous studies have suggested that field early type galaxies at high $z$ are more compact than those seen locally  
(e.g.,\citealt{Whitaker2012,Cassata2013,Williams2014} and references therein), though this work usually concentrates on the more 
luminous objects. Our above result on the surface brightness vs. luminosity plot (Fig.~\ref{subplot}) already suggests that we are not 
seeing particularly compact  luminous galaxies in our clusters.

However, in other measured parameters there are clear differences between the values for the high redshift sample and  those for 
the low redshift comparators. Red sequence galaxies in our high redshift clusters have, on average, lower S\'ersic indices, $n$,  than 
those in Virgo at the equivalent $H-$band magnitude (Fig.~\ref{figsiz}, middle panel) assuming passive evolution. We plot the distribution
of Sersic indices for all cluster galaxies and Virgo in the histogram (middle panel) in Fig.~\ref{figsiz}. This shows than $n$ is systematically
lower by about 1.5 compared to Virgo for our sample. This difference is confirmed by a K-S test that rejects the hypothesis that the two 
samples are drawn from the same parent distribution at more than $99.9\%$. 

Almost all the  Virgo red sequence galaxies down to an equivalent $H-$band magnitude of $M_H=-21$ have $n>3$ (typical of classical ellipticals) 
whereas most of the high redshift sources have $n<3$  (and usually $n<2$), excepting the most luminous and massive cluster galaxies. Even at an 
observed magnitude of $22<H<23$ ($-21.6<M_H<-20.6$) where there is increased contamination of the red sequence sample by interlopers, there 
are insufficient objects with high $n$ for the true red sequence galaxies to typically have the S\'ersic indices as high as their Virgo counterparts.  
The numbers of objects in the plots can be directly compared, as there are $\sim 50$ Virgo galaxies to the luminosities we consider, and about 
140 galaxies in total for the high redshift clusters. 

Similarly, we find that the high redshift sample contains more objects with flattened axis ratios (i.e., more disk-like) than in Virgo (bottom panels 
of Fig.~\ref{figsiz}). Again the histogram shows that the high redshift clusters have a long 'tail' at low $b/a$ and a K-S test also confirms that the 
two samples have not been drawn from the same distribution at more than the 5$\sigma$ level.The lack of objects in Virgo with $n < 2$ and 
$b/a < 0.5$ compared to the high redshift sample therefore appears to be significant, even accounting for some degree of contamination by 
interlopers. In all these figures, we plot all clusters at $z \sim 1.25$ compared to Virgo; no cluster is responsible for the observed 
results on its own and they all share the above features (i.e., a more disk-like and flattened population compared to the local data).

An obvious interpretation of this result is that most of the high redshift red sequence galaxies are  more disk-like or contain more significant 
disk components than their low redshift counterparts (in agreement with similar observations by \citealt{Rembold2012}). Only the brightest 
galaxies have S\'ersic indices consistent with those of genuine ellipticals, and even then they are usually lower than those of equivalently 
bright  and massive Virgo counterparts. Other parameters are consistent with this interpretation.  Clearly, many of the high redshift galaxies 
show very significant elongation, the entire distribution of axial ratios (Fig,~\ref{figsiz} bottom panel) is consistent with expectations from a 
sample of randomly oriented disks \citep[e.g.][]{Lambas1992}. 

Compared to the Virgo cluster, a population of more highly elongated objects is present at all luminosities, while such systems can only be seen 
for the fainter Virgo galaxies (and recall that even Virgo dwarf ellipticals, with $n \sim 1$ to 2, are round in shape; \citealt{Binggeli1995}).The range 
of axial ratios seen in the Virgo sample is much smaller with typically higher values. This is  consistent with an intrinsically elliptical/spheroidal population
implied by their higher S\'ersic indices.  Indeed, \cite{Chang2013} find an increased fraction of massive quiescent galaxies with flatter axial ratios at 
$z > 1$ even in the field \citep[see also][]{Bruce2012}.

\begin{figure}
\hspace{-0.5cm}\includegraphics[width=0.6\textwidth]{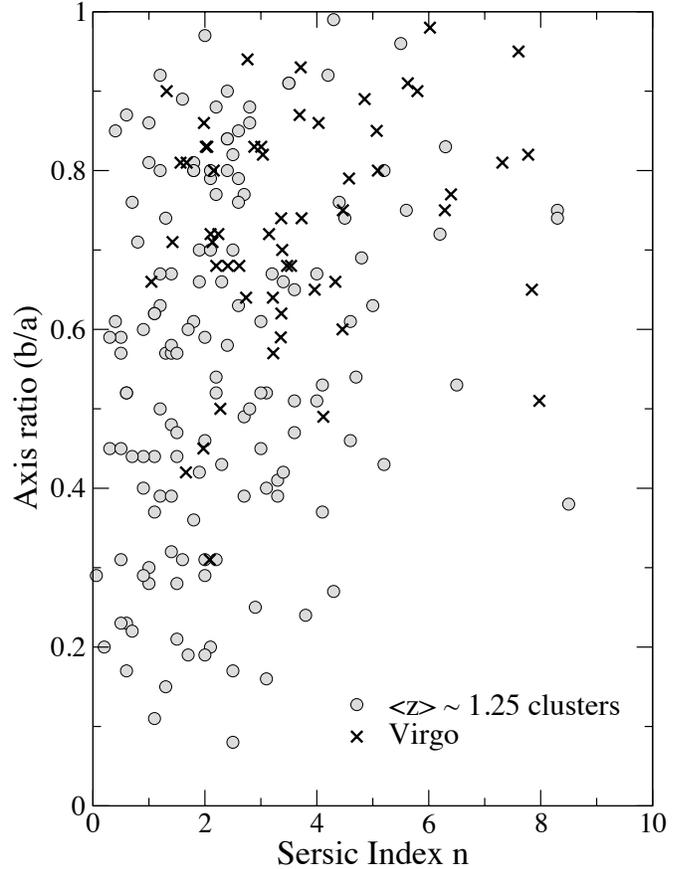}
\caption{S\'ersic indices of galaxies in our clusters vs. axis ratio, compared to data in the Virgo cluster from Ferrarese et al. (2006).
We see that at $z \sim 1.25$ clusters contain a population of low $n$ galaxies with flattened shapes, thus resembling disks.}
\label{figren}
\end{figure}

\begin{figure*}
\includegraphics[width=0.95\textwidth]{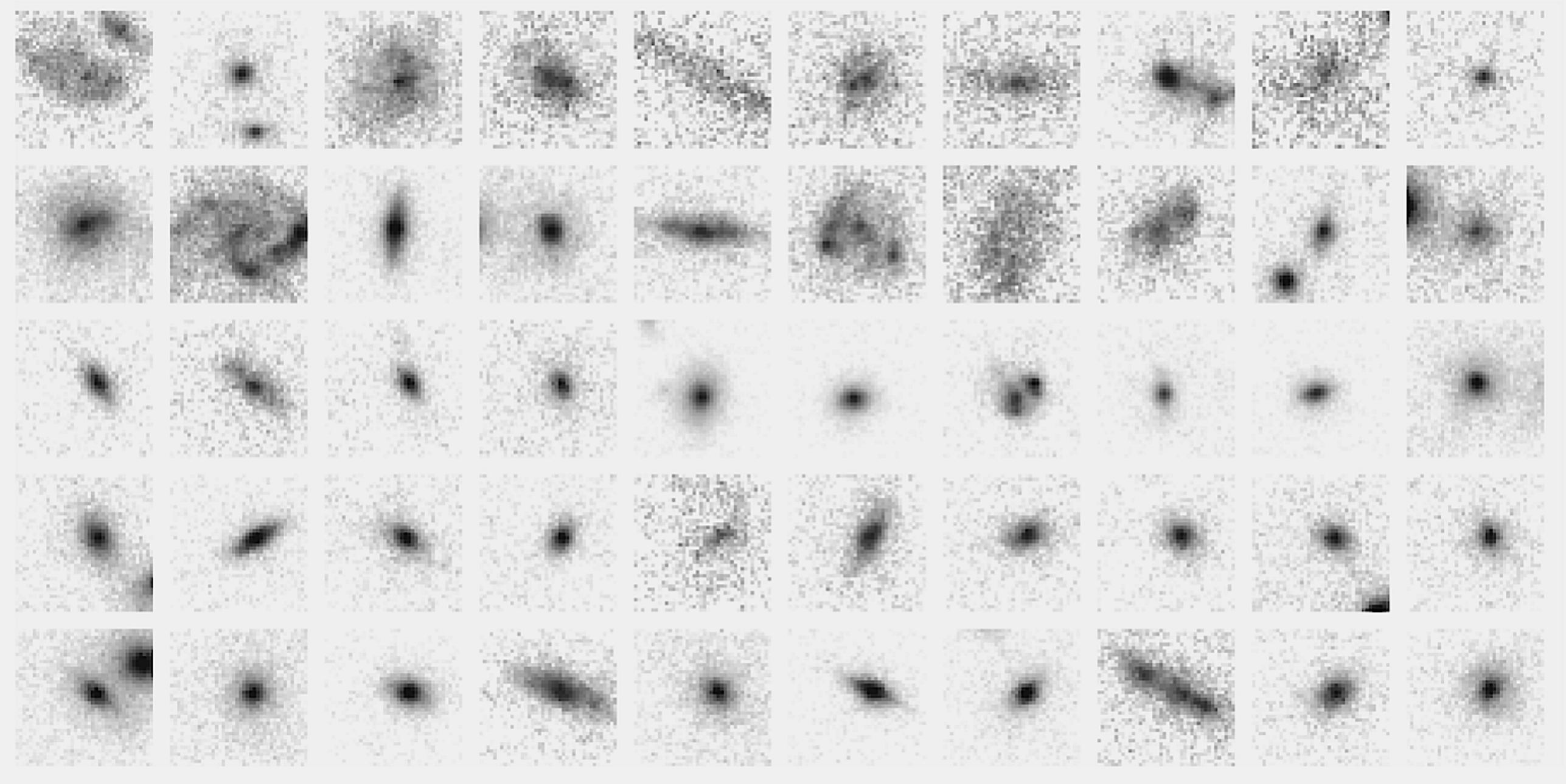}
\caption{HST/ACS $2\times2$ arcsec$^2$ $z-$band images for 50 red sequence galaxies with $n<2$ and magnitudes within  $\sim 1$ 
magnitude of $M^*$ drawn from all four clusters. Each image greyscale stretches between the mean sky level to the square-root of the 
central surface brightness (so more diffuse, lower surface brightness galaxies appear to have noisier images). }
\label{morph}
\end{figure*}

Fig.~\ref{figren} shows the same information in a different fashion, plotted against S\'ersic index, where we see that there are 
red sequence galaxies in high redshift clusters with significantly flatter axial ratios at lower $n$ than their Virgo counterparts. 
The more highly elliptical galaxies in Fig.~\ref{figsiz} have lower $n$ and correspond to the red sequence objects in the lower 
left region of  Fig.~\ref{figren}. All this points to a different (disky) morphology for most red sequence galaxies in our high $z$ clusters. 
Indeed, the range of projected $b/a$ suggests that these disks are more like the thin ones seen in modern day spiral galaxies, than 
the typically 'thicker' ones seen in SOs \citep{Laurikainen10}, the most common disk galaxies in low $z$ clusters. That is, cluster galaxies 
at this epoch  have become `red' without becoming `early type' yet, though they already have roughly the same stellar mass as their local 
counterparts. The images of the red sequence galaxies bear this out: Fig~\ref{morph} shows postage stamps of red sequence
galaxies (in the $z$ band) within $\sim 1$ mag. of $M^*$ and with S\'ersic indices $n<2$, demonstrating the disky nature of the bulk of 
the population.

\subsection{Colour gradients}

\begin{figure*}
\includegraphics[width=\textwidth]{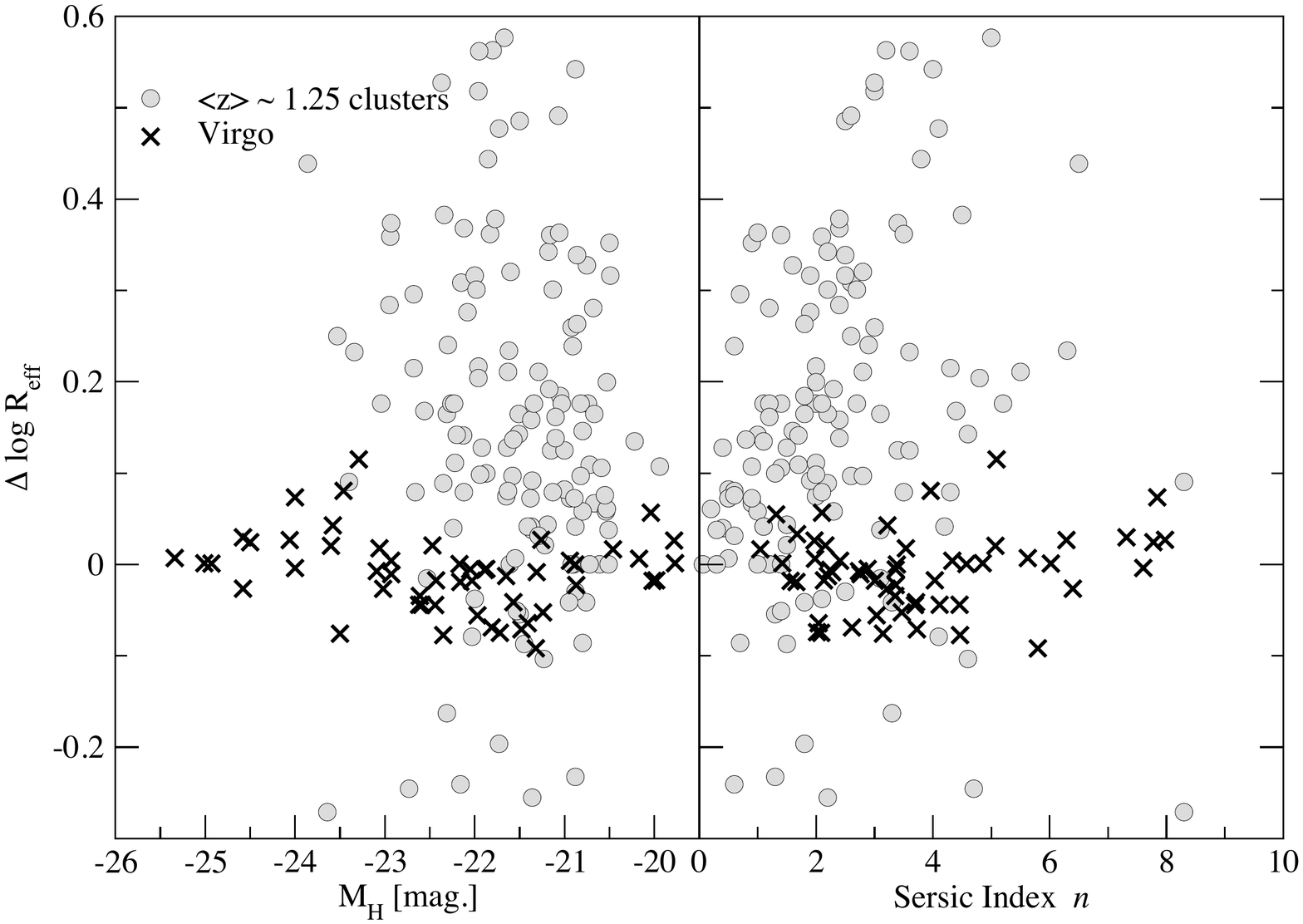}
\caption{Colour gradients of galaxies in our clusters (see legend) vs. absolute $H$ magnitude (left panel)
and S\'ersic index (right panel) compared to data in the Virgo cluster from Ferrarese et al. (2006). While nearly
all Virgo galaxies have small colour gradients, irrespective of luminosity or Sersic index, our galaxies have both
larger gradients and a wider spread, indicating the existence of age gradients with a large spread in ages for 
the outer regions of galaxies.}
\label{figgradre}
\end{figure*}

Local red sequence galaxies show weak negative (bluer outwards) colour gradients \citep{Tamura2000}; these weak gradients are 
observed to have the same size to $z \sim 0.6$, implying that they are due to metal abundance gradients \citep{LaBarbera2003}.
We measure the colour gradients in $z-H$ (approximately rest-frame $B-R$) for our $<z>=1.25$ galaxies following the method of 
\cite{LaBarbera2003}, i.e., we parameterise the gradients via the ratio of the effective radii as separately measured in the $z$ and 
$H$ bands, $\Delta \log R_{eff} = \log(R_{eff}(z)/R_{eff}(H))$. The reason we do this is that, for these faint and small objects, we cannot 
easily derive a radial luminosity profile which is accurate enough to derive colour gradients directly as in the lower redshift studies of
\cite{Vader1998} or \cite{Tamura2000}. For these high redshift systems, sky subtraction will be uncertain, the differences in the point 
spread functions between ACS and WFC3 are difficult to model and any small misalignment between the blue and red image may result 
in spurious results for directly determined colour profiles (e.g., from {\tt ellipse}). Our approach minimises these issues, as it simply compares 
the effective radii derived from each image and measures the colour gradient via their ratio. The centering of the profile, the effects of the PSF 
and sky level are modelled for each galaxy via GALFIT and are, in principle, accounted for.

As shown in Fig.~\ref{figgradre}, the high redshift galaxies are found to have significantly positive  $\Delta \log R_{eff}$ (i.e., a larger 
$R_{eff}$ in the bluer band), hence negative colour gradients in the usual sense in rest-frame $B-R$ (observed $z-H$). These are 
much larger than those observed for Virgo galaxies  (typically consistent with little or no gradient when measured in $g-z$). Here 
we find that  the high redshift  $ n < 2$ systems typically have somewhat larger $\Delta \log R_{eff}$  than found locally (0.1 to 0.4 as 
opposed to $<0.1$) while the high redshift  spheroids (large $n$) have $\Delta \log R_{eff}$ of $0.3$ to $0.5$. Using equation
(4) from \cite{LaBarbera2002}, we find that a median $\Delta \log R_{eff}$ of 0.13 corresponds to a colour gradient of $-0.25$ mag. 
in $B-R$ per decade in radius. Local systems in \cite{LaBarbera2002} have typical gradients of --0.1 to --0.3 mag., for comparison.
Similarly significant negative colour gradients in a substantial fraction of early type galaxies at $1.1 < z < 1.9$, albeit among field galaxies, 
have been measured by \cite{Gargiulo2012}.

This argues that these galaxies have significantly bluer exteriors than their local counterparts, while still being overall red in colour - star 
formation has likely ended throughout, but more recently in the outer parts.  Colour gradients in local red sequence galaxies  are
commonly attributed to a gradient in mean metal abundance with radius (e.g., Foster et al. 2009). The sizes of the colour gradients 
we measure are more consistent with age gradients, as they are too large to be produced by metallicity alone \citep{Saglia2000,
LaBarbera2003}, with higher redshift galaxies being younger at larger radii.  Star formation may therefore have continued for longer
periods in the outer parts of these galaxies, before they were quenched on to red sequence. (Of course, their descendants will also be
younger in the outer parts, but the colour differential decays rapidly with time). The large scatter in the colour gradients in Fig~\ref{figgradre}
suggests that cessation of star formation may have taken place over a range of times for each galaxy. This is similar
to the findings of \cite{Jorgensen2005}, where fainter cluster galaxies in a $z=0.83$ cluster were found to have had significant star
formation at $z \sim 1.5$ based on the tilt of its fundamental plane.

Considering our measurements of S\'ersic indices and axis ratios above, the most straightforward interpretation of these findings is one where, 
apart from the brightest ones, most red sequence galaxies in clusters at these redshifts  still possess significant disk components with younger 
mean ages for their stellar populations. The relative strength of the high redshift gradients can be explained assuming that the rate at which 
gradients decrease itself declines with increasing time since cessation of star formation in the disks (e.g., see Fig.~9 in \citealt{LaBarbera2003} 
for the evolution in colour with time of an age-induced gradient). Given that there is $\sim 3$Gyr between $z=0.6$ and $z=1.2$ and a similar 
time between $z=1.2$ and $z=3$  (the earliest that star formation is likely to end in any disk), it is unsurprising that  gradients of the observed 
strength are only seen at the higher redshifts studied in this work.

Given that the galaxies we are studying are red and have ceased significant star formation at an earlier epoch, the clear implication is that 
colour (stellar population) evolution precedes morphological evolution \citep{Skibba2009,Kovac2010} and that most of these red sequence galaxies  
are galaxies with significant disk components  that fade or evolve secularly into bulge-dominated red (quiescent) galaxies in the present universe 
(e.g., \citealt{Jaffe2011}). This is not the first time such an evolutionary scenario has been suggested to explain the properties of high redshift 
populations. \cite{Bundy2010} proposed a similar scheme for the evolution of red sequence galaxies at $1 < z < 2$ in the COSMOS field, 
where morphology changes {\it after} star formation is suppressed, as disks fade and bulge components become more significant (in a 
relative if not absolute sense). Although it is a field sample, several red disks are also observed by \cite{Bruce2012} in the CANDELS survey, 
albeit at somewhat higher redshifts.

Given that only a minority of the bright galaxies even in Virgo are true ellipticals, the majority being S0s \citep{Ferrarese2006}, 
it is likely that most of these ``red disks" will evolve into classical lenticulars rather than true ellipticals. While these high redshift galaxies 
could themselves be labelled as lenticulars (as we see no evidence of spiral arms), they are significantly diskier (in terms of their S\'ersic indices, 
range of axis ratios and colour gradients) than classical zero-redshift cluster S0s. Early on, \cite{Michard1994} proposed that all but the brightest
ellipticals can be observed to host disks and should be classified as S0s. With the exception of very luminous ellipticals, most
early-type galaxies are found to contain a rotating stellar disk \citep{Emsellem2011}, with several S0s found to be fast rotators
\citep{Emsellem2010}, and, by implication, therefore disky. In terms of the classification used by \cite{Bundy2010}, 
based on their S\'ersic indices, our sample of cluster red sequence galaxies  are mainly ``early disks" with true ellipticals only dominating at the 
high mass end. The stellar mass at which the dominant population changes is similar in the clusters studied here and in the highest redshift 
subsample of \cite{Bundy2010} (e.g. see their figure 2). 

While the red disky systems discussed here appear relatively extreme in their properties (S\'ersic indices, range of axis ratios and colour gradients), comparable red disk galaxies have been identified at lower redshifts. Passive red disks, with little or no spiral structure, have been found to be a notable sub-population in  the red sequence of the Abell 901/2 supercluster at $z=0.165$ \citep{Wolf2009}
\cite{Balogh2009} finds dust-reddened spirals with early-type morphology (S0/a) on the red sequence in groups, although these 
objects tend to be star-forming rather than passive. About 10\% of red sequence galaxies are observed to have infrared colours 
indicative of active or recent star formation in the PRIMUS sample \citep{Zhu2011}. On the other hand, \cite{Masters2010} identify 
truly passive spirals in their sample and show that they tend to have large masses, likely similar to our targets. These red disks have
similar, or slightly younger, ages than ellipticals \citep{Robaina2012,Tojeiro2013}, which is likely the case for our sample as well, 
given our interpretation of the observed colour gradients as age gradients.

Passive spirals are known in nearby clusters as well \citep{vandenBergh1976,Koopmann1998}. Their frequency may even increase 
in the higher redshift MORPHS sample \citep{Poggianti1999}, where they are believed to gradually replace S0s. These passive red
spirals also tend to occur more commonly at intermediate densities \citep{Bamford09, Masters2010} and although the numbers are small, many of our objects seem to lie somewhat outside of the cluster core. Similarly, \cite{Ferre2014} show that in CL0152-13 at $z=0.83$
a small population of cluster members lying outside of the two main subclusters has younger ages than the apparently old ellipticals
in the cluster core.

These red disks may originate from quenched disk galaxies, which are abundant in local groups \citep{Feldmann2011}. \cite{Carollo2013,Carollo2014} find that quenching and fading of disks may be responsible for the apparent evolution of the 
typical size of the early type galaxy population as a function of time. At higher redshift, \cite{Fontana2004} and \cite{Abraham2007} 
claim that at $z > 2$ many 'red' galaxies in the field are  disk-dominated. Intermediate mass disks are present at $z=1$ in the 
COSMOS field, but vanish by $z=0.2$ and it is likely that they form bulges by secular evolution \citep{Oesch2010} as the merger 
rate is too low to account for the observed decrease in their space densities. The velocity dispersion in the cluster environments 
of the high redshift red sequence galaxies also significantly limits the significance of mergers in their evolution, while our observations
also show that mergers must not have been important in these objects. This implies that whatever the mechanism for any disk fading/bulge growth that causes the population to evolve morphologically with time, it does not have to involve significant merger 
activity across a wide range of environments \citep{Vulcani2013}. \cite{Cibinel2013} find that the timescale for morphological evolution, 
by disk removal or fading, is $\sim 2-3$ Gyrs in their group samples. Based on the evolution of the colour gradients, that vanish at 
least by $z \sim 0.6$ \citep{LaBarbera2003}, the disks we observe at $z=1.25$ appear to disappear on similar timescales. Whether
this occurs by disk fading, removal or secular evolution is an issue that can  be addressed with better and more data on clusters above $z=1$ than present in the archive currently. Such a dataset would also allow us to considerably improve our estimate for the evolution 
of the luminosity function parameters.

\section{Conclusions}

In this work we have demonstrated that while the stellar populations of typical $z\sim1.25$ cluster red sequence galaxies appear to have already reached a state of  only passive evolution, their morphologies must clearly evolve over the subsequent  $\sim 8.5$~Gyr. Specifically we find the following:
\begin{itemize}

\item  Red sequence cluster galaxies in four $<z>=1.25$ clusters appear to have formed
their stars and assembled their mass completely by this redshift, at least down to 3 magnitudes
below the $M^*$ point, evolving passively afterwards.

\item Unlike field galaxies, these red sequence galaxies do not show size evolution when compared to 
zero redshift  cluster galaxies. 

\item However, there is clear evidence that these objects must continue to evolve {\it morphologically} 
at these redshifts. Apart from the most massive of the high redshift galaxies, these systems appear to have lower S\'ersic 
indices than those of a similar mass in low redshift clusters, projected axis ratios extending to lower (b/a) than local counterparts, 
and clear negative colour gradients much larger than those encountered locally. These colour gradients are so large that they can only really be attributed to  age gradients in the stellar populations, with the galaxies being younger outwards.  

\item Taken together all of these observations imply  that the bulk of the red sequence galaxies in massive $z\sim 1.25$ 
clusters are  galaxies with significant disk components (diskier than present-day lenticulars) unlike those identified in the dense regions of
low redshift clusters which have the characteristics of classical  ellipticals and S0s.

\item Given the clear difference in  the S\'ersic indices and colour gradients in the two epochs, a clear prediction 
is that the disks must reduce in prominence over time through ageing of their stellar populations and/or their 
secular evolution into bulges. The present results do not suffice to determine whether this is due to disk fading (implied
by the observation that the effective radii do not change), bulge luminosity growth (suggested by the increase in the S\'ersic index) or both.

\end{itemize}

\section*{Acknowledgments}

We would like to thank Inger J{\o}rgensen and Anna Cibinel for useful discussions. This work is based on observations made with the NASA/ESA Hubble Space Telescope, and obtained from the Hubble Legacy Archive, which is a collaboration between the Space Telescope Science Institute (STScI/NASA), the Space Telescope European Coordinating Facility (ST-ECF/ESA) and the Canadian Astronomy Data Centre (CADC/NRC/CSA). Some of the data presented in this paper were obtained from the Mikulski Archive for Space Telescopes (MAST). STScI is operated by the Association of Universities for Research in Astronomy, Inc., under NASA contract NAS5-26555.
The PIs of the original projects which produced
these data are thanked for providing excellent deep archival
images with diverse uses beyond the original programmes. We also thank the anonymous referee for a helpful report that helped to clarify a number
of issues.

\bsp

\label{lastpage}

\end{document}